\documentclass[twocolumn,superscriptaddress,nofootinbib]{revtex4-1}

\usepackage{graphicx,xcolor,amsmath}
\usepackage{hyperref}
\usepackage{bm}
\usepackage{mathtools}
\usepackage{booktabs}

\usepackage{tikz}
\usepackage{tkz-euclide}
\usetikzlibrary{backgrounds}
\usetikzlibrary{decorations.pathmorphing}	
\tikzset{
    v/.style={decorate, decoration={snake, segment length=3mm, amplitude=0.75mm}, draw},
    v2/.style={decorate, decoration={snake, segment length=3mm, amplitude=0.5mm}, draw},
    f/.style={draw=black, postaction={decorate},
        decoration={markings,mark=at position .6 with {\arrow[very thick]{latex}}}},
    fb/.style={draw=black, postaction={decorate},
        decoration={markings,mark=at position .4 with {\arrowreversed[very thick]{latex}}}},
    fnar/.style={draw=black},
    g/.style={decorate, draw=black,
        decoration={coil,amplitude=3pt, segment length=3.5pt}},
    s/.style={dashed,draw=black, postaction={decorate},
        decoration={markings,mark=at position .55 with {\arrow[very thick]{latex}}}},
    sb/.style={dashed,draw=black, postaction={decorate},
        decoration={markings,mark=at position .55 with {\arrowreversed[draw=black,very thick]{latex}}}},
    snar/.style={dashed,draw=black,line width =1.25pt},
}
\DeclareRobustCommand\mymixing{\tikz[anchor=base, baseline]{\draw[line width=1] (0,0.08) circle (0.1cm);\node[] at (0,0){$ \times $};}}
\usetikzlibrary{shapes.misc}
\usepackage{tikzsymbols}

\definecolor{mypurple}{RGB}{164,64,214}

\newcounter{qnumber}

\begin{document}

\title{Light vectors coupled to bosonic currents}

\author{Jeff A.\ Dror}
\email{jdror@lbl.gov}
\affiliation{Department of Physics, University of California, Berkeley, CA 94720, USA}
\affiliation{Ernest Orlando Lawrence Berkeley National Laboratory, University of California, Berkeley, CA 94720, USA}
\author{Robert Lasenby}
\email{rlasenby@stanford.edu}
\affiliation{Stanford Institute for Theoretical Physics, Stanford University, Stanford, CA 94305, USA}
\author{Maxim Pospelov}
\email{mpospelov@perimeterinstitute.ca}
\affiliation{Perimeter Institute for Theoretical Physics, 31 Caroline Street N, Waterloo, Ontario N2L 2Y5, Canada}
\affiliation{Department of Physics and Astronomy, University of Victoria, Victoria, BC V8P 5C2, Canada} 
\date{\today}

\begin{abstract}
	New spin-1 particles with masses below the weak scale are present
	in many theories of beyond Standard Model (SM) physics.
	In this work, we extend previous analyses by systematically considering
	the couplings of such a vector to the bosonic sector of the SM, focusing
	on models that lead to mass-mixing with the $Z$ boson.
	These couplings generically lead to enhanced emission of the vector's longitudinal
	mode, both in Higgs decays and in flavor changing meson decays. We present bounds in the SM+$ X $ effective theory and investigate their model-dependence. For the case of Higgs decays, we point out that tree-level
	vector emission is, depending on the model, not always enhanced, affecting
	the constraints.
	For meson decays, which are the dominant constraints
	at small vector masses, we find that while $B$ decay constraints can be weakened by fine-tuning UV parameters, it is generically difficult to suppress the stringent constraints from kaon decays.
\end{abstract}

\maketitle

\section{Introduction}

Light new states, with masses below the weak scale, arise in many
theories of physics beyond the Standard Model (SM). These may have
gone undetected if their couplings to the SM are sufficiently weak.
If such particles are not associated with many other new states
at the same mass scale, then they should generically
have spin $\le 1$; here, we will concentrate on new light vector
bosons, which have been extensively studied as mediators to a dark
sector~\cite{Boehm:2003hm,Pospelov:2007mp,ArkaniHamed:2008qn},
and as explanations for experimental
anomalies~\cite{Gninenko:2001hx,Kahn:2007ru,Pospelov:2008zw,TuckerSmith:2010ra,Batell:2011qq,Feng:2016ysn}.

The lowest-dimensional interaction between a new vector and SM states
is the dimension-4 coupling to a SM current, $X_\mu J^\mu_{\rm SM}$.
Unless $J_{\rm SM}$ is conserved, including quantum effects, the SM +
$X$ effective theory is non-renormalizable. This manifests 
as $({\rm energy}/m_X)$-enhanced amplitudes involving the longitudinal mode of $X$, 
for processes with $\partial_\mu J^\mu \neq 0$.
Such processes can be the dominant $X$ production mechanism at high-energy
colliders, mostly importantly through emission
in flavor changing neutral current (FCNC) $B$ and $K$ decays.

In previous work~\cite{Dror:2017ehi,Dror:2017nsg}, we systematically
considered the the phenomenological
consequences of general couplings to SM fermions, $X_\mu J^\mu_{{\rm SM},f}$.
The only fully conserved currents in the SM are the electromagnetic (EM) current,
and (if neutrinos are Dirac) the $B-L$ current; any other coupling will give rise
to enhanced longitudinal emission in some circumstances
(though if non-conservation is only through neutrino masses,
e.g. for $L_\mu - L_\tau$, this will be very suppressed). Various models have been considered in the literature; 
light vectors having a chiral anomaly with the SM~\cite{Dror:2017ehi,Dror:2017nsg,Ismail:2017fgq} (see
also~\cite{Ismail:2017ulg,Fox:2018ldq} for
vectors above the weak scale),
vectors with axial couplings~\cite{Fayet:2006sp,Fayet:2007ua,Kahn:2007ru,Barger:2011mt,Karshenboim:2014tka,Dror:2017nsg},
and ones with weak-isospin
violation~\cite{Karshenboim:2014tka,Dror:2017nsg}.

Here, we extend our previous analyses by considering light vectors with
couplings to bosonic SM fields. In particular, a $XW^+W^-$ coupling will
contribute to emission in FCNC penguin amplitudes, leading to similar
phenomenology as for couplings to fermions~\footnote{We
 can write the $W^+W^-$ current as a sum of the EM current and
a fermionic current,
showing that longitudinal emission is the same as for the fermionic current.}\cite{Dror:2017nsg}. New
phenomenology can also arise from a coupling to the SM Higgs, leading to
exotic Higgs decays~.
Such effects have been considered in a number of previous papers~\cite{Davoudiasl:2012ag,Davoudiasl:2013aya,Davoudiasl:2015bua};
in this work, we consider the possibilities more systematically,
investigating how signatures vary across different models.

For calculational simplicity, we will mostly work with UV completions
that do not introduce extra electroweak symmetry breaking (EWSB).
This means that, in the SM + $X$ effective field theory (EFT), $X$ must couple to a SM-gauge-invariant
current (or via Wess-Zumino terms, which are determined by the SM
fermion couplings of $X$~\cite{Dror:2017nsg,Dror:2017ehi}).
The lowest-dimension couplings with non-trivial effects
are
\begin{align}
  J^\mu_{\rm SM,b} & = \partial_\nu B^{\mu\nu} ;~\partial^\mu ( H^\dagger H);   ~H ^\dagger i\overleftrightarrow{ D }^\mu H;   ~H ^\dagger i \overleftrightarrow{ D} _X ^\mu H \,. \label{eq:start} 
\end{align}
where $D^\mu$ corresponds to the SM covariant derivative and $ D ^\mu _X $ includes a U(1)$ _X $ contribution, $ D ^\mu _X   \equiv D ^\mu + i g _X X ^\mu / 2 $.
The first current in (\ref{eq:start}) is the familiar ``kinetic mixing"
operator \cite{Holdom:1985ag}, and upon the use of equations of motion
for $B_\mu$ can be replaced with the hypercharge current $J^\mu_{Y}$.
The $X_\mu \partial^\mu ( H^\dagger H)$ coupling is equivalent, on integration
by parts, to $-(\partial^\mu X_\mu) H^\dagger H$, and replacing $\partial^\mu X_\mu$
via the equations of motion gives a higher-dimensional operator.

The $H ^\dagger  D _{(X)} H$ currents
are of more interest
(the subscript $ ( X ) $ is used to denote either the SM or SM+U(1)$ _X $). In unitary gauge, they lead to a mass mixing of $X$ with the $Z$ boson,
\begin{equation} 
{\cal L} \supset  g_X X_\mu\big( H ^\dagger i\overleftrightarrow{ D }^\mu _{ ( X ) } H \big)  ~\longrightarrow ~\varepsilon  _Z m _Z ^2  X _\mu Z ^\mu \,. \label{eq:deltam}
\end{equation} 
Transforming to the mass basis, this results in the light vector
state gaining a coupling to the SM neutral current. In particular, this
gives a coupling to $W^+ W^-$, resulting in FCNC decays, and
a coupling to the Higgs, resulting in exotic Higgs decays. Consequently,
the mass mixing coupling provides an excellent prototype for
the phenomenology of enhanced longitudinal emission.
This coupling has also been the subject of
several phenomenological studies (see {\em e.g.,}\cite{Davoudiasl:2012ag,Lee:2013fda,Davoudiasl:2013aya,Davoudiasl:2014kua,Davoudiasl:2015bua,Kim:2014ana,Kong:2014jwa,Davoudiasl:2014mqa,Xu:2015wja,Izaguirre:2015eya,DelleRose:2017xil,Campos:2017dgc,Abdullah:2018ykz,Bertuzzo:2018itn}), including motivations for new parity-violation experiments
\cite{Davoudiasl:2015bua,Souder:2015mlu,Safronova:2017xyt}. 

Previous studies of 
$X$-$Z$ mixing have generally assumed that it leads to 
$1/m_X$-enhanced amplitudes for $h \rightarrow ZX$ at tree level~\cite{Davoudiasl:2012ag,Davoudiasl:2013aya,Davoudiasl:2015bua}.
This is true if $\varepsilon  _Z m _Z ^2  X _\mu Z ^\mu$ is the
only operator in the EFT,
but not if it arises from the SM gauge invariant operators above.
If we were to ignore the SM Yukawa couplings,
then 
$H ^\dagger D _{(X)} H$ would be conserved within the SM.\footnote{This is true for both currents for single $ X $ emission, though $H ^\dagger  D  H$ is not conserved
in processes involving multiple $X$s, with consequences, for example, in $h \rightarrow XX$.}
Consequently, the tree-level $h \rightarrow XZ$ amplitude
is \emph{suppressed} at small $m_X$.
To obtain significant $1/m_X$-enhanced amplitudes, we need to consider processes
involving heavy fermions. Examples include FCNC penguin amplitudes, and
Higgs decays through a top loop, which we will study in detail.

The $h \rightarrow ZX$ amplitude is finite at one loop,
so the rate for this processes is predicted in the EFT, up
to corrections suppressed by the scale of heavier states.
However, FCNC penguin amplitudes (and $h \rightarrow XX$)
are divergent at one loop. Hence, in the limit that the scale
of new states $\Lambda$ is significantly higher than the EW scale, the dominant
contribution to the amplitude will scale as $\log (\Lambda^2 / m_{\rm EW}^2)$.
For SM-gauge-invariant $X$ couplings, the coefficient of this term is 
determined within the EFT.
There will also be threshold corrections from the high-scale physics, which 
are not suppressed by inverse powers of $\Lambda$.
Both of these effects mean that, without knowledge of the UV physics, only
a generic prediction can be made for FCNC amplitudes.

If $\Lambda$ is not too high, then the log-enhanced term may be
numerically comparable to other contributions. In particular, one might
worry that there could be cancellations between this piece and other
contributions, significantly altering the rates for meson decays. In
this work, we illustrate the robustness of FCNC constraints within a
two Higgs doublet model (2HDM) model, though we expect our conclusions to
hold more generally. While $B \rightarrow K X$ rates can be suppressed
by tuning the 2HDM parameters (in particular, the mass of the charged
Higgs), $K^\pm \rightarrow \pi^\pm X$ amplitudes can only be suppressed
by a factor of a few, unless the model has additional flavor structure.

The $1/m_X$-enhanced amplitudes under discussion arise from the
non-renormalizability of the SM + $X$ EFT, which must have a cutoff at
(or below) a scale parametrically set by $m_X/g_X$. Standard UV completions,
such as the 2HDM examples given later, result in higher-dimensional EFT operators
involving $X$ being suppressed by $g_X$, as well as by inverse powers of the cutoff
scale. Hence, the effects of dimension $> 4$ operators are less important.

\section{Further Comments on bosonic current portal}
\label{sec:ewc}
The $H ^\dagger D_{(X)} H$ couplings
have the following unitary gauge decomposition:
\begin{align}
g_X X_\mu H^\dagger  i\overleftrightarrow{D} ^\mu  H \label{eq:higgs} 
		& =   m _Z ^2  \left(1 + \frac{h}{v}\right)^2 \varepsilon _Z X_\mu Z^\mu\,, \\
g_X X_\mu H^\dagger  i\overleftrightarrow{D} _X ^\mu  H  & =  m_Z^2 \left(1 + \frac{h}{v}\right)^2 \left( \varepsilon _Z  Z_\mu + \varepsilon _Z ^2 X _\mu \right)  X^\mu \,. \label{eq:higgs2}
\end{align} 
where $v \simeq 246 ~\rm{GeV}$ is the electroweak vacuum expectation value (VEV), $h$ 
is the physical (125\,GeV) Higgs boson, and $ \varepsilon _Z \equiv g _X v / m _Z $. The additional contribution in $ D _X $ leads to terms proportional to $X _\mu X ^\mu $, resulting in the current being conserved even in processes involving multiple $X$s (but still no SM Yukawa insertions). As mentioned above, this form is somewhat
deceiving as one may expect it to lead to unsuppressed $ h X _\mu X ^\mu
$ and $ h X _\mu Z ^\mu $ couplings. To see this is not the case we work
in the mass basis where we parametrize the couplings as,
\begin{align} 
  {\cal L} & \supset  C _{ hXZ}  \left( \frac{ h }{ v } + \frac{1}{2}  \frac{ h ^2 }{ v ^2 }\right) X _\mu Z ^\mu \notag \\ & \qquad + \frac{1  }{2} C _{ h XX}\left(  \frac{ h }{ v }  + \frac{1}{2} \frac{ h ^2 }{ v ^2 } \right) X _\mu X ^\mu  \,, \label{eq:ZXcoup}
\end{align}  
where the relation between the Higgs and Higgs-squared couplings is guaranteed by assuming the direction of electroweak symmetry breaking is aligned with the Higgs. We find,
\begin{align} 
&   C _{ h XZ} =   -  2 \varepsilon _Z m _X ^2 \,,  C _{ h XX} = 2 \varepsilon _Z ^2 m _X ^4/  m _Z ^2   &  &  (H ^\dagger D _X   H )  \notag \\ 
&   C _{ h XZ} =   -  2 \varepsilon _Z m _X ^2 \,, C _{ h XX} =  2 \varepsilon _Z ^2 m _Z ^2   &&    (H ^\dagger D  H ) \,.
\label{eq:cval}
\end{align} 

We conclude that the ${\cal M} _{h\to ZX} \propto m_X^{-1}$ is not a generic consequence of 
$ZX$ mass mixing, as it cancels for the simplest 
SM gauge invariant realization. Notice, however, that if such mixing is realized as part of a more complicated current, {\em e.g.} 
$ X_\mu(H^\dagger  i\overleftrightarrow{D} ^\mu  H )(H^\dagger   H ) \supset Z_\mu X^\mu $, $m_X^{-1}$ behavior can appear even at tree level. This is because $(H^\dagger  i\overleftrightarrow{D} ^\mu  H )(H^\dagger   H )$ is not conserved in the purely bosonic sector. 
Similarly, if there are additional sources of EWSB, then this spoils the
conservation of the U(1)$ _X $ current and can also lead to enhanced
rates (as discussed below in the case of a 2HDM model).

While important for Higgs physics, these couplings with $h$ will not affect the one-loop results for the FCNC amplitudes. 
Therefore,  the mass mixing $X_\mu Z^\mu$ operator and full SM gauge-invariant couplings, 
$X \cdot H^\dagger  D _{ ( X ) }    H $, at leading loop level will give identical answers for FCNCs (as will any model having
the same $XW^+W^-$ and $X$-quark vertices).  
Nevertheless, the SM gauge invariant theory is more convenient for calculations,
as it allows these to performed in arbitrary gauges.
Keeping the Goldstone modes in the Lagrangian, in addition to the couplings in \eqref{eq:higgs} or \eqref{eq:higgs2}, we have,
\begin{equation}
	{\cal L} \supset \frac{1}{2} g_X X^\mu \left(G^+ i\partial_\mu G^-
	-  g v \,G^+ W^-_\mu + \dots \right) + {\rm h.c.} 
\end{equation}
where the ellipses denote neutral component terms which will not be relevant for us here.

It is instructive to discuss possible UV completions of (\ref{eq:higgs}) and (\ref{eq:higgs2}). If the breaking of U(1)$_X$ occurs at high energy scale
$\Lambda$, then vector portal under the discussion is itself an effective operator, 
\begin{align}
\frac{1}{\Lambda^2} \left(\Phi^*  i\overleftrightarrow{D} ^\mu  \Phi\right)\left( H^\dagger  i\overleftrightarrow{D} ^\mu  H\right) \label{eq:reduction} 
		\to   g_X X_\mu \left( H^\dagger  i\overleftrightarrow{D} ^\mu  H\right),
	\end{align}
where $\Phi$ is the `dark' Higgs of U(1)$_X$, and $D_\mu$ between $\Phi^*$ and $\Phi$ is the covariant derivative 
with respect to U(1)$_X$, $D^\mu = \partial^\mu +i g' X^\mu$, so that the coupling $X$ used previously relates to the U(1)$_X$ charge $g'$ as 
$g_X = g' |\langle \Phi\rangle |^2/\Lambda^2$. Such couplings of $X$ were discussed in {\em e.g.} \cite{Fox:2011qd}. While explicitly 
and separately SM and U(1)$_X$ invariant, operator (\ref{eq:reduction})
is of dimension 6, and needs further UV completion. A straightforward
UV completion can be built using sets of heavy vector-like fermions: $ N _{ {\rm SM}} $ (charged only under SM), $ N _{ X} $ (charged only under U(1)$ _X $), and $ N $ (singlet under all gauge symmetries). The charges are assigned
in such a way that $\Phi NN_X$ and $HNN_{\rm SM} $ Yukawa
interactions are allowed. This way, one fermionic loop of $N-N_X-N_{\rm SM}$ mixed propagator 
creates (\ref{eq:reduction}), and the effective energy scale that would resolve this interaction is linked to the fermion masses. 

A UV completion of (\ref{eq:higgs2}) can be accomplished with a 2HDM model as considered in~\cite{Davoudiasl:2014kua}.
This introduces two $SU(2)_L$ doublets $H_1$ and $H_2$;
the first is `SM-like', coupling to SM fermions ($ \psi _i $),
while the second has no fermion couplings, but couples to $X$.
Schematically,
\begin{equation}
	{\cal L} \supset D_\mu H_1^\dagger D^\mu H_1
	+ y _{ij} H_1 \psi _{i} \psi _{j}
	+ D_{X} ^\mu  H_2^\dagger D _{X,\mu }  H_2\,,
\end{equation}
where the covariant derivative acting on $H_2$ contains a U(1)$_X$
contribution. To avoid introducing extra EWSB into the SM + $X$ EFT, requires the light scalar Higgs to be aligned with the direction of EWSB. In the notation of~\cite{Branco:2011iw}, this is the `decoupling limit'
$\cos(\alpha - \beta) = 0$, where $\alpha$ is the angle between the neutral
components of $H_2,H_1$ and the scalar mass states, and $\beta$
is the angle between $H_1,H_2$ and the direction of EWSB. Then, writing
$H_2 = s_\beta H + c_\beta \tilde{H}$, where $H$ is the SM Higgs doublet
and $\tilde{H}$ consists of heavier states that do not obtain a VEV, we can integrate
out $\tilde{H}$ to obtain an effective coupling of the form  of~(\ref{eq:higgs2}),
\begin{equation} 
 (D_{X,\mu} H_2)^\dagger (D^\mu _X  H_2) \supset \underbrace{- \frac{  g _X }{ g / 2c _W } s _\beta ^2} _{ \varepsilon _Z }  X _\mu  H ^\dagger i \overleftrightarrow{ D} _{ X} ^\mu H \,.
\end{equation} 
The mass of $X$ in this model receives two independent contributions;
one from the VEV of $H$, and another from a Stueckelberg mass term in the
Lagrangian.
In more general 2HDM models, where $h$ is not aligned with the direction
of EWSB, we obtain couplings beyond (\ref{eq:higgs}) and (\ref{eq:higgs2}),
allowing enhanced Higgs decay rates at tree level~\cite{Davoudiasl:2012ag,Davoudiasl:2013aya,Davoudiasl:2015bua}.

Lastly, one might wonder if its possible to remove the dangerous $ m _X ^{-1} $-enhanced terms by charging additional fermions under U(1)$ _X $, resulting in a conserved current. The simplest way of doing it is suggested by the conservation of the hypercharge current, that can be split into bosonic and fermionic parts:
\begin{equation} 
0=\partial_\mu J^\mu_Y = \partial_\mu\big( H ^\dagger i\overleftrightarrow{ D} ^\mu H  
 +\sum _{i \in \rm SM } Y _i \bar{\psi} _i \gamma ^\mu \psi _i \big).
\end{equation} 
Thus, if the bosonic Higgs current is completed by the addition of the fermionic hypercharge current, one would remove $m_X^{-1}$-enhanced terms. This would correspond to the ``usual'' case of the kinetic mixing/hypercharge portal. 

\section{Fermionic processes}

SM fermion masses result in non-conservation of the axial part
of the neutral current,  
so the enhancement      
of longitudinal $X$ emission will occur the most in processes       
involving heavy fermions. For example, $X$-emission in the top          
quark production and decay, $gg(q\bar q)\to t\bar t X $ or $t\to        
WbX$, will occur with the rate proportional to $\varepsilon _Z^2        
m_t^2/m_X^2$~\cite{Kong:2014jwa,Kim:2014ana}. Consequently, should this 
parameter be large, both the top quark production cross section and its 
decay width will be affected. Given $O(5-10\%)$ accuracy in measuring   
the inclusive rates, setting very strong bounds on $\varepsilon _Z$     
does not seem realistic. A somewhat better sensitivity could likely be  
achieved employing exclusive searches of light $X$ final states.        

The most important constraints originating from the $X$-exchange are
from $ \nu _e -e $ scattering, and atomic parity violation (APV). Neutrino
scattering and APV are insensitive to the longitudinal mode of $X$, and
at very small $ m _X $ the constraints are $m_X$-independent. The most powerful limits from $ \nu _e - e
$ scattering are from a combination of the TEXONO~\cite{Deniz:2009mu,Li:2002pn,Wong:2006nx,Chen:2014dsa}, LSND~\cite{Auerbach:2001wg}, BOREXINO~\cite{Bellini:2011rx}, GEMMA~\cite{Beda:2009kx}, and CHARM II~\cite{Vilain:1993kd} experiments. Ref.~\cite{Bilmis:2015lja} summarized the limits for a $ B - L $ vector and we expect similar limits to apply for a mass mixed vector with the identification, $g _{ B - L} \leftrightarrow e \varepsilon _Z $. Since these are not constraining in any of our parameter space we include these here using this rough approximation.

Atomic parity violation and electron parity violation
scattering experiments have long considered a mass-mixed vector as a
prime target for their analyses~\cite{Benesch:2014bas,Aoki:2017ugb}.
We directly implement their limits here as described
in~\cite{Porsev:2009pr,Bouchiat:2004sp}.

\section{Higgs decays}

The simplest processes involving bosonic couplings of $X$
are the exotic Higgs decays $h \rightarrow Z X$ and
$h \rightarrow XX$, possible for $m_X < m _h - m _Z $ 
and $m_X < m_h/2$ respectively. As per the discussion in
previous sections, if the UV completion does not introduce extra
EWSB, then the $h \rightarrow ZX$ rate is not enhanced as $m_X \rightarrow 0$,
at least at tree level.
For the $H^\dagger i D^\mu_X H$ coupling, the $h \rightarrow XX$ rate is likewise unenhanced at tree level. 
However, as we calculate in this section, loop-level effects from heavy fermions
restore the small-$m_X$ enhancement.

\begin{figure} 
\begin{center} 
  \includegraphics[width=8cm]{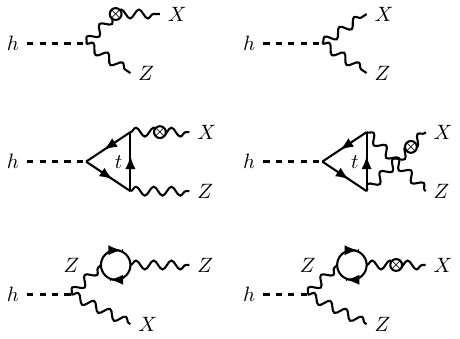}
 \end{center}
\caption{Leading contributions to $ h \rightarrow ZX $ decay at one-loop, where the ``\mymixing'' denotes a mass mixing.
At tree level there is a cancellation between the mixing term and
the direct $ h ZX $ coupling. Similar diagrams are present for $ h
\rightarrow XX $ decays.} 
\label{fig:hdecay}
\end{figure}

At one loop the effective coupling is given by the diagrams shown in Fig.~\ref{fig:hdecay}.
Note that in principle one should also include mass and field renormalization corrections on the Higgs, however these corrections all lead to corrections suppressed by extra powers of $ m _X $ and we drop them here.  The effective operators are straightforward to compute with only the axial coupling contributions leading to operators of the form in~\ref{eq:ZXcoup}.

After the inclusion of counterterms for the masses and the mixing parameter the divergence in the $ h X _\mu Z ^\mu $ is canceled leaving behind a finite piece given by,
\begin{align} 
C _{ h ZX} ^{ {\rm loop}} & =   \varepsilon _Z   \frac{ N _c g _Z ^2  m _t ^2 }{ ( 4\pi )  ^2  } \,, 
\end{align} 
where $ g _Z \equiv e / s _W c _W $.
These should be added with the tree-level couplings in~\ref{eq:cval} in either model to get the total couplings. Note that this expression has an opposite sign to the tree level contribution and hence will interfere destructively when the two are comparable (at larger $ m _X $).

Contrary to $ h Z _\mu X ^\mu $, the $ h X _\mu X ^\mu $ effective vertex is log-divergent and hence dependent on the UV completion. The effective vertex takes the form,
\begin{equation} 
C _{ hXX} ^{ {\rm loop}} = \varepsilon _Z ^2 \frac{ N _c g _Z ^2 m _t ^2 }{ (4\pi)^2 } \log \frac{ \Lambda ^2   }{ m _t  ^2 } \,.
\end{equation} 
These give the decay rates,
\begin{align} 
\Gamma  _{  h \rightarrow ZX } & = \frac{\left|  C _{ hZX} \right| ^2 m _h }{ 16 \pi v ^2  m _X ^2    }   \left[ 2w + \frac{ ( 1 - w - z ) ^2 }{ 4 z } \right]  \sqrt{   \lambda  ( 1 , w , z )} \,,\\ 
\Gamma _{  h \rightarrow 2 X } & =   \frac{  \left| C _{ hXX} \right| ^2m _h ^3  }{ 128\pi v ^2 m _X ^4  }   \left[ 1 - 4 w + 12 w ^2 \right] \sqrt{ \lambda ( 1 ,w,w )   }\,,
\end{align} 
where $ \lambda ( 1 , w, z )  \equiv  1 + w ^2 + z ^2 - 2 z w - 2 w - 2 z  $ and $ w \equiv m _X ^2 / m _h ^2 $, $ z \equiv m _Z ^2 / m _h ^2 $. Note that these are of the same form as those computed at tree-level (using equation~\ref{eq:cval}) but involve $ C _{ i} $ which are not suppressed by powers of $ m _X $. As promised, the (loop level) decay rates are always enhanced and scale as $ m _X ^{ - 2 } $ and $ m _X ^{ - 4} $ respectively.

Such production of $ X $ has been searched for by ATLAS~\cite{Aad:2015sva,Aaboud:2018fvk}. The $ h \rightarrow ZX $ only applies for a limited range of masses, $  m _h - m _Z > m _{X } > 15 ~\rm{GeV} $, due to the large $ h \rightarrow Z Z ^\ast $ background leading to a requirement of reconstructing the invariant mass of the decay products of $ X $. On the other hand, $ h \rightarrow XX $ applies in a broader mass range, $ {\rm GeV} \lesssim  m _X < m _h / 2 $ though involves more powers of $ \varepsilon _Z $ leading to slower improvement with luminosity. We recast the limits here in Figure~\ref{fig:ez} with the solid lines corresponding to couplings to the $ H ^\dagger D _X  H $ current and the dotted ones for coupling to $ H ^\dagger D H $. 

Interestingly we find that rare Higgs decays are already a competitive constraint in the high mass range. As expected the constraints are stronger for the $ H ^\dagger D H $ current, already surpassing atomic parity violation. Further we note that limits for $ h \rightarrow ZX $ drop at lower masses due to a cancellation between the loop and tree level contributions. If experimental thresholds could be lowered then such constraints would also be enhanced at smaller $ m _X $. Lastly we note that the solid line Higgs decay constraints also apply on the 2HDM of~\cite{Davoudiasl:2012ag} in the alignment limit, with the new loop induced constraints becoming most important at lower masses. With more data we expect future searches for rare Higgs decays will be the most constraining for heavier vectors also coupled to the more-evasive $ H ^\dagger D _X H $ current. 

\section{FCNC amplitudes with longitudinal $X$ emission}
In this section we evaluate the amplitudes for FCNC decays via longitudinal $X$ emission, using the SM + $X$ EFT. The SM flavor changing meson decay amplitudes are suppressed by the weak scale (schematically denoted, $ m _{ {\rm EW}}  $), whereas on-shell $ X $ amplitudes may only be suppressed by the $X$ coupling. This can lead to FCNC decays via an on-shell longitudinal $X$ enhanced by $(m _{ {\rm EW}} / m_X)^2$ relative to their SM counterparts, partially making up for the $g_X^2$ suppression.
This effect, combined with the large statistics and high precision achieved in experimental studies of 
$K$ and $B$ decays (as opposed to {\em e.g.} Higgs boson decays)
make them a powerful tool in constraining $\{m_X,g_X\}$ parameter space. 

\begin{figure}
  \begin{center} 
\includegraphics[width=8cm]{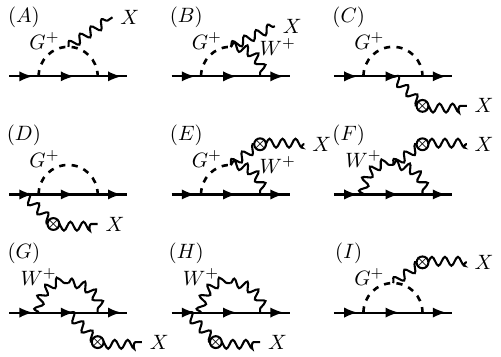} 
\end{center}
  \caption{$bsX$ amplitude from the SM+$ X $ EFT for $ X $ coupled to either the $ H ^\dagger  D   H $ or $ H ^\dagger  D  _X H $ current. The ``\mymixing'' denotes a mass mixing between the $ X $ and $ Z $. Diagrams (B) and (E) include swapping the $ G ^+ $ and $ W ^+ $ lines while ( D ) and (H) include emitting the $ Z $ from the outgoing fermion lines.}
\label{fig:eftdiag}
\end{figure}

Evaluating the down-type FCNC amplitude, $d_i d_j X$, in the SM + $X$ EFT,
involves the sum of diagrams in Figure~\ref{fig:eftdiag} (as a consequence of not including additional sources of EWSB there is no $  XW ^+ W ^-  $ vertex in this basis).
Diagrams (A) and (B) involve the $X G^+ G^-$ and $XG ^+ W ^- $ couplings,
while (C)-(I) come from the $X_\mu Z^\mu$ mixing.
In an $R(\xi)$ gauge, the amplitude from (C)-(I)
is, at leading order in external momenta, simply $\varepsilon _Z$ times
the SM $Z$-penguin result~\cite{Inami:1980fz}, which is finite. However,
the charged Goldstone coupling in (A) results in an uncanceled 
logarithmic divergence. Writing the total amplitude in terms
of a low-energy effective interaction, we have
\begin{align} 
 {\cal L}  & \supset  g _{ d _i d _j X} \bar{d } _i \gamma _\mu P _L d _j X ^\mu  \,, \\ 
 g _{ d _i d _j X} &\simeq   \frac{ g ^3  }{  2 c _W } \frac{\varepsilon  _Z}{ 16 \pi ^2 } 
	\sum_{q \in \{u, c, t\}} V_{qd_i} V_{qd_j}^* f(m_q^2/m_W^2)\, ,
		\label{eq:veft}
\end{align} 
where $f$ depends on the UV parameters.
Assuming a wide scale separation, and the dominance of the logarithmic contribution, in the EFT one finds: 
\begin{equation}
\label{eq:EFT}
	f(x) = - \frac{x}{4} \log \frac{\Lambda^2}{m_W^2} +  c_{\rm th},
\end{equation}		
where $c_{\rm th}$, associated with threshold corrections, is sub-leading to the logarithmic part
in the limit where the high scale, $ \Lambda \gg m_W$. As usual, these threshold corrections come from both the IR scales (associated here with the 
weak scale), and the UV scale. Strictly speaking, it is only within a specific UV completion 
that one could fully identify the scale $\Lambda$, and define $c_{\rm th}$.
In particular, while the coefficient of the log-enhanced term
in the EFT calculation is gauge-independent, the other parts of the amplitude
are not, illustrating that they depend on the UV completion.
If additional states are at a sufficiently high scale, then we would
expect the log-enhance term to dominate allowing concrete constraints to be set. Conversely, for smaller $ \Lambda $, threshold corrections can be significant opening the possibility of canceling it against the log-divergent piece. We discuss this more quantitatively in an explicit UV completion in Section~\ref{sec:thdm}.

In unitary gauge, only the $Z$-mixing diagrams contribute,
and the amplitude is just given by $\varepsilon _Z$ multiplied by the SM
$Z$-penguin. This may be slightly confusing, since amplitudes
for physical processes in the SM must be finite. 
Indeed, for the loop-induced FCNC decay of an on-shell $Z$-boson, $Z\to d_i\bar d_j$, 
such divergence is absent.  The explanation is that
the $Z$ mass in the SM is related to its couplings,
in a way that makes the divergences in the external-momenta-dependent terms
in the FCNC amplitude (which we neglected above) exactly
cancel the momentum-independent divergences~\cite{He:2009rz}.
For SM FCNC amplitudes involving off-shell $Z$ exchange, 
such as $d_i\to d_j l \bar l$, divergences in the effective $d_i-d_j-Z$
vertex  cancel against the divergences from 
$W$ box diagrams~\cite{Inami:1980fz,He:2009rz}. However, $m_X$ and $g_X$ are independent
parameters, so this cancellation will not occur for
FCNC decays via an on-shell $X$. Consequently, computing in unitary
gauge, where diagram (A) is absent, also gives rise to divergences in the SM + $X$ EFT,
resulting in the same $\log \Lambda^2 / m_W^2$ contribution
as the $R(\xi)$ gauge calculation.

While the top-quark loop will in general dominate, it is also useful to determine the 
EFT answer for the light quark contributions. 
If all of the up-type quarks had the same mass, then CKM unitarity would
mean that the sum in equation~(\ref{eq:veft}) cancels --- at least, that the
EFT-calculable parts do. A UV completion with additional flavor
structure could lead to the $c_{\rm th}$ contributions varying between
different up-type quarks. However, 
for a UV completion without such structure, without loss of generality we can take $f(x) \rightarrow 0$
as $x \rightarrow 0$, since all up-types quarks being massless
would lead to zero total amplitude.
Within the EFT, $f(x) \propto x \log x$ for small $x$.
This means that even for $s-d-X$ vertex, the internal up and charm contributions are always subdominant
to the internal top one (unless special cancellations occur), despite the corresponding CKM elements
 being larger.

\begin{figure*}
  \begin{center} 
\includegraphics[width=1.5\columnwidth]{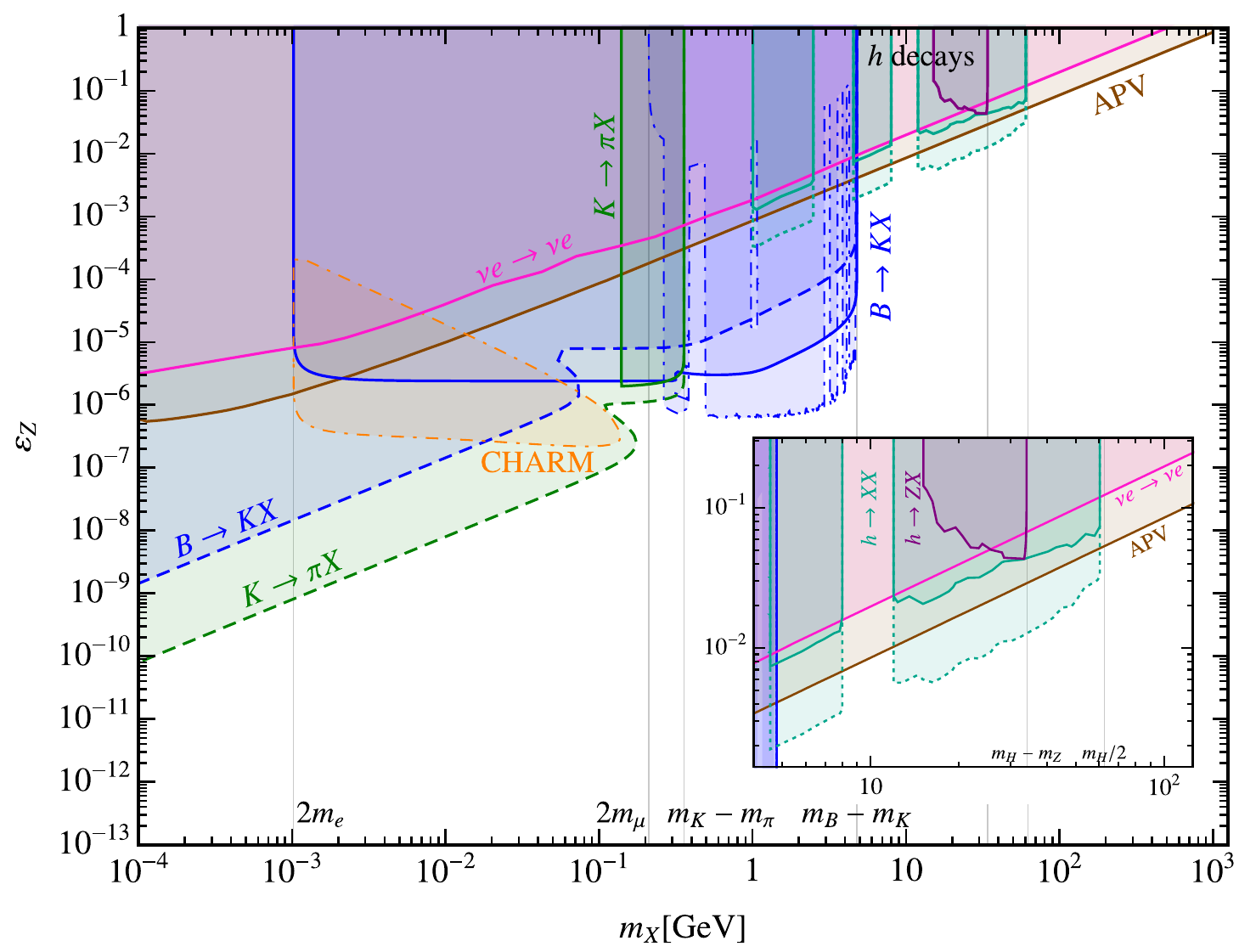} 
\end{center}
	\caption{Constraints on a vector $X$ with a `mass-mixing' coupling
	$\varepsilon _Z m_Z^2 Z_\mu X^\mu$ to the SM $Z$ boson. The 
	FCNC rates from such a coupling are UV dependent
	--- the degree to which this
	can affect constraints is discussed in the text.
We use the limits on $ B $ decays
from~\cite{Grygier:2017tzo,Aubert:2008ps,Aaij:2016qsm},
from~\cite{Artamonov:2008qb,AlaviHarati:2003mr} for $ K $ decays
(\cite{Crivellin:2017gks} for SM prediction for $ K \rightarrow \pi
\bar{\nu} \nu $ decay), from~\cite{Bergsma:1985qz} for the CHARM
beam dump, from~\cite{Bilmis:2015lja} for $ \nu -e $ scattering, and finally from~\cite{Porsev:2009pr,Bouchiat:2004sp} for
computing the constraints on atomic parity violation. The Higgs decay constraints arise from direct searches for $ h \rightarrow XX $ and $ h \rightarrow ZX $ with the solid color correspond to a vector coupled to $ H ^\dagger D  _X  H $ and the dotted to a coupling to $ H ^\dagger D H $~\cite{Aaboud:2018fvk}.}
\label{fig:ez}
\end{figure*}

To summarize, FCNC amplitudes in the SM + $X$ EFT run logarithmically,
so are dependent on the UV physics. However, their generic scale
is set by the $X$ couplings in the EFT. The same structure applies
for $X$ couplings to SM fermions, as analyzed in~\cite{Dror:2017nsg}.
For sufficiently light $X$, all of these lead to FCNC meson decays
with rates enhanced by $\sim m_{\rm EW}^2 / m_X^2$ compared to competing
SM processes (this is for down-type FCNCs --- up-type FCNC rates are suppressed by at least additional powers of the charm mass over the weak scale).
As discussed in~\cite{Dror:2017nsg}, the corresponding $B$ and $K$ meson decays can provide strong
constraints on non-renormalizable $X$-SM couplings.
Figure~\ref{fig:ez} illustrates these constraints for
a `typical' UV model by taking $\Lambda \sim {\rm TeV}$ in equation (\ref{eq:EFT}),
and including only the log-enhanced term.
In the next Section, we will
see how this relates to the results in an example UV theory.
The figure also shows other constraints arising from the SM couplings
of $X$, including neutrino-electron scattering and 
parity violation experiments, illustrating that when FCNC
constraints apply, they are generically significantly stronger than
other probes.

As mentioned in the introduction, any model that results
in a $XW^+W^-$ vertex (and/or couplings to quarks) will
generically contribute to FCNCs in a similar way.
The $\varepsilon _Z$ model here is illustrative in that it results
in UV-divergent and $m_t^2/m_W^2$ enhanced amplitudes
(in contrast to e.g.\ Wess-Zumino couplings, which lead to finite
FCNC amplitudes~\cite{Dror:2017ehi,Dror:2017nsg}).
As in the case of couplings to SM fermions~\cite{Dror:2017nsg},
this is the generic and strongest behavior.

\subsection{Beam dump constraints}
In our previous work~\cite{Dror:2017ehi,Dror:2017nsg}, we stated that
$K \rightarrow \pi X_L$ decays would be the dominant $X$ production mechanism
in proton beam dump experiments such as CHARM~\cite{Bergsma:1985qz}, following the analysis in~\cite{Dolan:2014ska}.
However, the details of this analysis are incorrect, since almost all of the kaons
produced in such experiments will be stopped in the target before decaying.\footnote{We
thank Felix Kahlhoefer for bringing this issue to our attention.}

The consequences of kaon interactions in the target are investigated
in~\cite{Winkler:2018qyg}, which adopts the conservative procedure of
estimating the number of kaons that decay before their first interaction.
Considering $X$ production from only these kaon decays results in a significantly
smaller yield than we used in~\cite{Dror:2017ehi,Dror:2017nsg}.
However,
it is still the case that $K \rightarrow \pi X_L$ decays are only suppressed
by $m_{\rm EW}^2 / (m_X^2/g_X^2)$ compared to SM channels. On the other hand,
the production rate from `$\pi^0$-like' interactions, as estimated
in~\cite{Bergsma:1985qz, Essig:2010gu}, is suppressed by $\sim f_\pi^2 / (m_X^2/g_X^2)$
compared to the SM $\pi^0$ production rate.
Consequently, even taking the
the small kaon survival rate from~\cite{Winkler:2018qyg}, $X$ production from 
kaon decays still dominates over the $\pi^0$ estimate. In addition to $ X $ produced from kaon decays, there is a subdominant component arising from production of $ B $ mesons, which decay into vectors with significantly larger boosts~\cite{Dobrich:2018jyi}. However, due to the smaller primary yield of $B$ mesons (using the estimate in~\cite{Winkler:2018qyg}) $X$ production from kaons, when kinematically allowed, is typically larger. Our constraints from the CHARM beam
dump experiment (plotted in Figures~\ref{fig:ez} and~\ref{fig:ezmh}), which use the kaon and B efficinecy estimates from~\cite{Winkler:2018qyg}, illustrate
that beam dumps can still constrain new parameter space.

Neutrino beam experiments, in which pions and kaons are not stopped
in the target, but instead focussed into a collimated beam, result in
a much larger kaon decay yield, and would consequently give more $X$
production. However, the detectors in such experiments are generally set
up to detect neutrino scattering events, which would constitute
a background to the $X$ decays we want to search for. In contrast,
the beam-dump mode run of CHARM~\cite{Bergsma:1985qz}
used a large, air-filled detector volume, yielding a basically background-free
experiment (this difference is illustrated in~\cite{Bergsma:1983rt}).


\section{2HDM example}
\label{sec:thdm}

The SM + $X$ EFT calculation in the previous
section gives the `generic' scale of FCNC amplitudes.
However, for UV completions in which new physics is not too far above the EW scale, it is possible that other contributions
will be numerically comparable to the log-enhanced term.
In particular, one might wonder about the robustness of limits for particular UV completions,
as partial cancellations in the amplitude may weaken
FCNC constraints.
In this Section, we will perform a full calculation of
FCNC amplitudes (to leading order in external momenta) within
a simple UV theory, illustrating these points.

\begin{figure} 
\centering 
\includegraphics[width=\columnwidth]{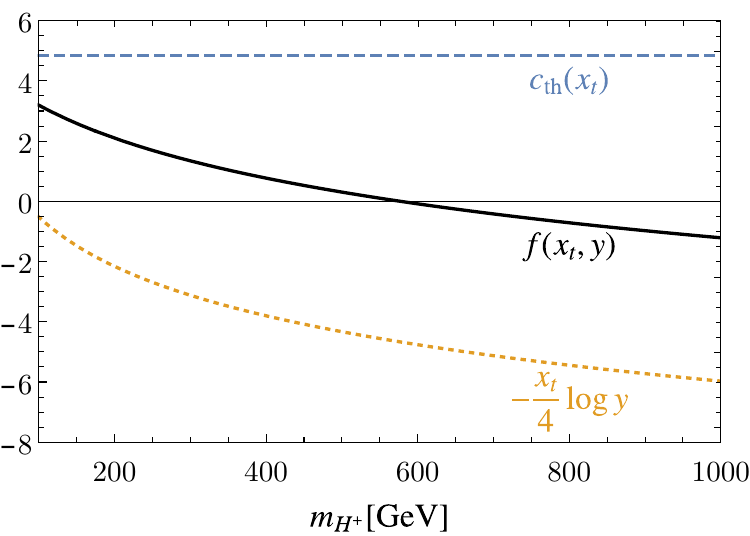}
\caption{Different contributions to the effective $ d _i d _j X $ coupling from top loops as given in equation~\eqref{eq:fequation} (with $ x _t \equiv  m _t ^2 / m _W ^2 $). The dashed blue and dotted orange lines denote the threshold and large-logarithm pieces respectively while the black line denotes the full $ 1 $-loop function. We see a cancellation between the threshold and running contributions around $ m _{H ^+ } \simeq 575 ~\rm{GeV}$ beyond which the coupling is well approximated by the log-enhanced piece.}
\label{fig:thdm}
\end{figure}

As our reference model we use the 2HDM summarized in section~\ref{sec:ewc}.
Calculating FCNC vertices in the full 2HDM theory, the UV divergences
coming from EW penguin diagrams are canceled by diagrams involving
charged Higgs exchange (Figure~\ref{fig:thdm}). Computing these,
we find that the total contribution to the FCNC amplitude is given by
\begin{equation}
	g_{d_i d_j X} \simeq -\frac{g_X g^2 s_\beta^2}{16 \pi^2}
	\sum V_{q d_i} V_{q d_j}^* f \Big(\frac{m_q^2}{m_W^2}, \frac{m_{H^{+}}^2}{m_W^2}\Big)\,,
\end{equation}
where
\begin{align} 
f ( x , y ) & \equiv \frac{ x }{ 4 } \bigg\{ 2 + \frac{ y }{ y - x } - \frac{ 3 }{ x - 1 } + \frac{ 3 ( y - 2 x + 1 ) }{ ( y - 1 ) ( x - 1 ) ^2 } \log x \notag \\ 
& \qquad +  \frac{ y ( y ^2 - 7 y + 6 x ) }{  (y - 1  )(  y - x  ) ^2 } \log \frac{ x }{ y }  \bigg\}\,,
\label{eq:fequation}
\end{align} 
and we use $ x _q \equiv m _q ^2 / m _W ^2 $ and $ y \equiv m _{H ^+ }  ^2 / m _W ^2 $.

For $y$ large, $f(x,y) \simeq - ( x / 4 )  \log y$, in agreement with the EFT result in equation~(\ref{eq:veft}).
Identifying the scale $\Lambda$ with $m_{H^+}$, we also find the threshold correction in 
the limit of large $\Lambda$,
\begin{equation}
c_{\rm th} = \frac{x (3 x^2 - 9 x + 6 + (x^2 - 2 x+4 )\log x )}{4 (x - 1 )^2}\,.
   \label{eq:cth}
\end{equation}
It is easy to see that for $x$ given by the top mass, the sign of $c_{\rm th}$ is opposite to $-\frac{1}{4} x \log y$. 
For small enough $x$, we have $f \simeq x \log x$, 
again as expected from the EFT. Our results are in agreement with results in previous studies~\cite{Davoudiasl:2014kua} which computed the amplitude by relating the theory to that of an axion with fermionic couplings~\cite{Freytsis:2009ct}. As discussed in the previous section, the top-quark contribution is typically dominant, for both $K$ and $B$ decays. 

\begin{figure}
  \begin{center} 
\includegraphics[width=\columnwidth]{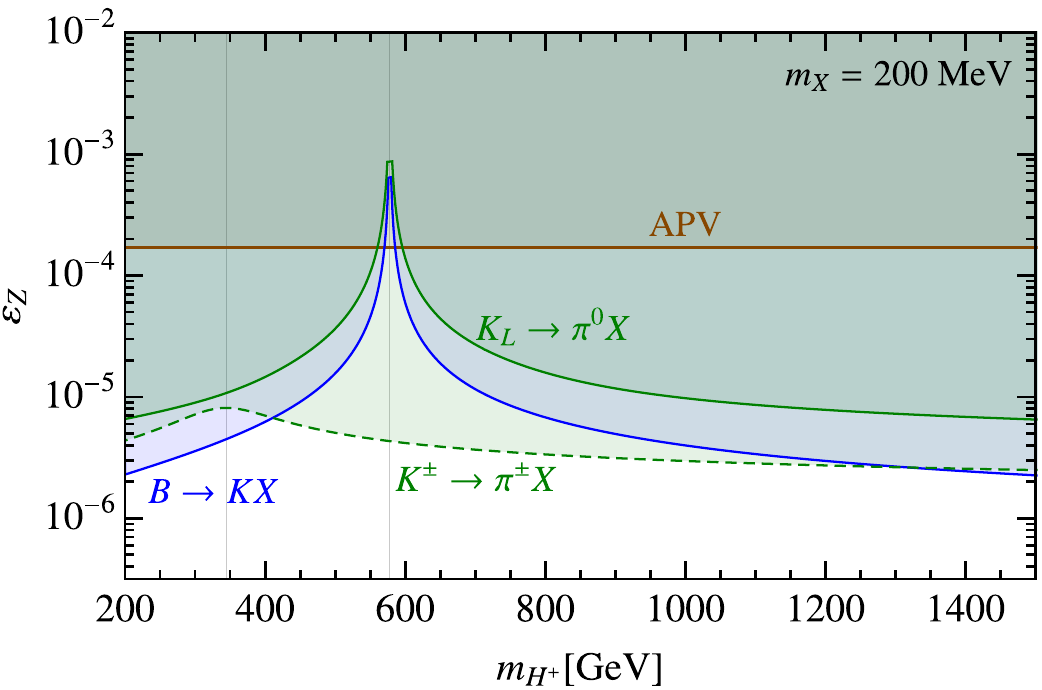} 
\end{center}
\caption{Constraints on $\varepsilon _Z$ in the 2HDM model of
	Section~\ref{sec:thdm}, for a light vector mass of
	$m_X = 200~{\rm MeV}$. As discussed in the text,
	varying the UV physics (here, the charged Higgs mass)
	changes the FCNC amplitudes involving $X$, and consequently
	changes FCNC meson decay rates. For $K^+ \rightarrow \pi^+ X$
	decays, the impact is mild, whereas for $B \rightarrow K X$ and $K_L \rightarrow \pi^0 X$ decays,
	there is a specific charged Higgs mass for which the rates are significantly
	reduced. Also plotted are atomic parity violation constraints~\cite{Porsev:2009pr,Bouchiat:2004sp} on $\varepsilon _Z$, which
	do not depend on the UV completion --- these are always weaker
	than $K^+$ decay constraints, and can only compete
	with the other meson decays in a narrow $m_{H^+}$ range. 
	}
\label{fig:ezmh}
\end{figure}

The different threshold and running contributions are compared with the full expression in Fig.~\ref{fig:thdm} showing a partial cancellation in the flavor changing rate at $ m _{ H ^+ } \simeq 575 ~{\rm GeV} $ for the top contribution. Note, however, that since there are additional contributions to the total FCNC decay amplitude that have a different complex phase to the top quark contribution, the total amplitude does not pass through zero as we change $m_{H^+}$. For kaon decays,the next most important contribution to the $sdX$ vertex is from the charm-quark contribution. The argument\footnote{We use the Particle Data Group parametrization of the Cabibbo-Kobayashi-Maskawa matrix~\cite{Tanabashi:2018oca}.} of 
$V_{cs} V_{cd}^* \simeq -0.22 + 10^{-4} i$ is significantly smaller than that of $V_{ts} V_{td}^* \simeq 10^{-4} (-3.3 - 1.4 i)$, and to large extent this mitigates the small charm mass, $f ( x _c ,y ) /f ( x _t , y )  \propto {\cal O} (m_c^2/m_t^2)$.
Consequently, as illustrated in Figure~\ref{fig:ezmh}, the limits from $K^\pm
\rightarrow \pi^\pm X$ decays can only be relaxed by a factor of a few 
by tuning $m_{H^+}$.
On the other hand, for $B$ decays, the charm-quark CKM product $V_{cb} V_{cs}^*
\simeq 0.04 + 10^{-6} i $ has similar magnitude
to the top-quark product, $V_{tb} V_{ts}^* \simeq -0.04 + 7 \times 10^{-4} i$, and 
the charm loop will not play an important role
(the cancellation will instead mostly be lifted by external momentum corrections,
which are suppressed by $\sim m_b^2/m_W^2$ compared to the top
contribution).

Figure~\ref{fig:ezmh} illustrates how these cancellations affect the experimental
constraints on the coupling of the new vector, for a representative
value of $m_X = 200~{\rm MeV} $ (where both $K$ and $B$ decay constraints apply).
As noted above, for `generic' UV completions, the FCNC
constraints are significantly stronger than other experimental probes.
The figure shows that $B \rightarrow K X$ and $K_L \rightarrow \pi^0 X$ constraints
(the latter process is CP-violating, so only depends on the imaginary part of the amplitude, which generically goes through zero) can be significantly relaxed at the same
value of $m_{H^+}$. However, the $K^\pm \rightarrow \pi^\pm X$ constraints
are not significantly canceled around that point; overall, the limit on $g_X$
varies only by a factor $\sim 2$ across the entire range of $m_{H^+}$.
Nevertheless, the reduction of the log contributions due to threshold corrections is clearly an important effect. 
While constraints in Fig.~\ref{fig:ez} are plotted using the simplified form $f (  x _t , y )  = -(x_t/4)\times \log({\rm TeV}^2/m_W^2)$, 
the same value of $f ( x , y ) $ in the UV complete version is achieved with $m_{H^+} \simeq 7.5 ~{\rm TeV} $. 

While these calculations were performed in the context of a particular
UV completion, the generic point that different components
of the amplitude have different complex phases can be seen within the EFT.
Consequently, if FCNCs in the UV completion are dominantly controlled
by one parameter ($m_{H^+}$ in our 2HDM example), then varying
this parameter will generically not result in the amplitude
passing through zero.
Thus, unless the UV completion has additional flavor structure that enables
it to arrange such a cancellation, and/or multiple parameters which
affect the different-phase components in different ways, FCNC amplitudes can generically only
be canceled by a similar degree to that derived here.

Another point illustrated by this UV completion is that the scale
of new physics has to be sufficiently low for the partial cancellations to occur. 
For the case of extended Higgs sectors, the current constraints
on the mass of the new charged Higgs states range from
$ \sim 200 ~\rm{GeV}-600
~\rm{GeV} $ depending on the details of the Higgs sector
(see~\cite{Khachatryan:2015qxa,Khachatryan:2015uua,Sirunyan:2017sbn}
for CMS searches and~\cite{Aad:2015typ,Aad:2015nfa,Aaboud:2016dig}
for corresponding ATLAS searches).
These limits are statistics dominated, and should improve with the high
luminosity LHC run. In the considered UV completion, the most relaxed bounds on $g_X$ 
are for $m_{H^+}$ of 575 GeV, which is within reach of the LHC.


\section{Kinetic mixing and FCNCs}

The only fully renormalizable coupling of a new light vector
to the SM is via `kinetic mixing' with the hypercharge gauge
boson,
\begin{equation}
	{\cal L} \supset \frac{1}{2} \varepsilon _Y X_{\mu\nu} B^{\mu\nu}
\end{equation}
(if neutrinos are Dirac, then a coupling to the $B-L$ current is also
renormalizable). Defining electromagnetic and weak neutral currents of the SM 
as ${\cal L} \supset A_\mu J_{\rm EM}^\mu + Z_\mu J_Z^\mu$, to first order in the small
mixing $\varepsilon _Y$, we can perform field redefinitions so that
the new light spin-1 state has couplings
\begin{equation}
\label{eq:AZ}
{\cal L} \supset	\varepsilon _Y X_\mu \left(J_{\rm EM}^\mu \cos\theta_W + \frac{m_X^2}{m_Z^2} J_Z^\mu \sin\theta_W\right)\,,
\end{equation}
where $\theta_W$ is the weak mixing angle. 
While a coupling to the (non-conserved) neutral current leads to
$\mbox{energy}/m_X$-enhanced longitudinal amplitudes, the $m_X^2/m_Z^2$
suppression of this coupling means that such amplitudes
are suppressed by $m_X E / m_Z^2$ overall, and hence vanish
for $m_X \rightarrow 0$ as required by renormalizability.

Most treatments of `dark photons' --- new vectors 
with a coupling to the EM current in the low-energy (sub-EW-scale) SM
--- do not consider the small neutral current coupling 
that would arise from hypercharge mixing.
One reason is that, in non-renormalizable SM + $X$ models,
these couplings are not necessarily linked (e.g.\ 
one could always add a kinetic mixing to the models
in the previous section). Another is that, while the neutral current coupling
does give some experimentally distinct signatures,
the $m_X^2/m_Z^2$ suppression means that, for $m_X \ll m_Z$,
these are generally less important than the EM coupling signatures.

This point is illustrated by FCNC meson decays, which,
as discussed in previous sections, are generally the strongest
probes of a light vector's neutral current coupling.
For kaon decays,
the $K^+ \rightarrow \pi^+ \gamma^*$ amplitude can be computed
in chiral perturbation theory, using information on the $K\to 3\pi$ amplitude as input~\cite{DAmbrosio:1998gur}.
This gives the FCNC amplitude from the EM current coupling
of $X$ \cite{Pospelov:2008zw}.
The amplitude from the neutral current couplings is given
by the usual EW penguin contribution.
Due to the $m_X^2$ suppression of the neutral current coupling in (\ref{eq:AZ}),
both of these are $\propto m_X^2$.
However, the penguin contribution is suppressed by $V_{ts} V_{td}^*$,
which has magnitude $\sim 4 \times 10^{-4}$, whereas the
$\gamma^*$ amplitude is only suppressed by the Cabibbo angle.
Hence, the neutral current contribution to these decays
is subdominant to the `dark photon' contribution.
The same is true for $K_S \rightarrow \pi X$ decays.
For $K_L \rightarrow \pi^0 X$ decays, which are CP-violating,
the neutral current term dominates: however, the experimental
limits on these decays are comparable to those
for $K^+ \rightarrow \pi^+ X$, so the latter will generally be
a better probe of a hypercharge-mixed vector.
In the end, it turns out that the kaon decays do not put any additional restrictions on 
dark photons compared to standard ``bump hunt" searches, with the exception of models with
large $X\to invisible$ channel, where the competing SM rate, $K^+\to \pi^+ \nu\bar\nu$
has a very small branching. 

$B \rightarrow K X$ constraints on the kinetic mixing are even less important than the kaon constraints. 
$B \rightarrow K \gamma$ decay occurs with the branching ratio of $\simeq 4\times 10^{-5}$, and the corresponding decay to a ``dark photon"
will be smaller by a factor of $(\varepsilon_Y \cos\theta_W)^2$.
Current experimental constraints require $\varepsilon_Y \lesssim 10^{-3}$ over
the relevant mass range, which renders the $B$ decays well outside
experimental capabilities.
Contributions of neutral currents do not change this conclusion.


\section{Conclusions}

We have discussed some of the simplest ways of couplings new light
vectors to the bosonic sector of the SM, and their phenomenological
consequences. In particular, we have focused on the lowest-dimensional,
SM-gauge-invariant currents to which a new vector can couple. Besides
kinetic mixing, the only two non-trivial currents are $H ^\dagger D H
$ and $ H ^\dagger D _X H $. Coupling a light vector, $X_\mu$, to these
currents leads to an effective $Z-X$ mass mixing, as well as couplings
of $X$ to the scalar Higgs. The most observationally significant effects
of these couplings come about through $m_X^{-1}$-enhanced longitudinal
$X$ emission, analogously to couplings to non-conserved fermionic
currents~\cite{Dror:2017nsg}.

An important property of the $H ^\dagger D _{ ( X ) } H $ currents is
that they are conserved in the purely bosonic sector of the SM for
single-$ X $ emissions. Hence, one should not expect an enhancement
of the longitudinal $X$ emission at tree-level. Nevertheless, this
property is broken by quantum corrections and $m_X ^{-1} $-behavior
is present when fermions are included, as the Yukawa/mass terms break
current conservation. An important places to look for the source of the
constraints are the loop-induced processes which can profit from the
high energies available in the virtual states. Loops of top quarks can
induce important signals for both currents of interest and we highlight
their potential to probe such vectors.

For sufficiently light vectors, flavor changing decays of the $K$
and $B$ mesons are extremely sensitive probes of bosonic couplings,
imposing more stringent constraints than parity violation
or scattering experiments.
Applying the EFT calculation to the SM+$ X $ theory, we compute the FCNC
amplitudes, identifying the leading $ \log(\Lambda/m_W)$ behavior.
To explore the relation of the effective theory to its UV completions
and the possibility of a cancellation eliminating these constraints,
we study a two Higgs doublet UV completion in detail, with one of
the Higgs fields charged under U(1)$_X$. We find that the threshold
corrections, as it is often the case, contribute with the opposite
sign to the logarithmically enhanced terms. At $m_{H^+} \sim 575$\,GeV
there is nearly perfect cancellation of the $bsX$ amplitude. We argue,
however, that such cancellation is not universally applicable to all
enhanced $d_id_jX$ amplitudes. Up to the caveat of an exotic case of UV
completion with additional flavor structure, FCNC contributions with
different complex phases cannot generically be simultaneously canceled,
leading to the conclusion that $K^+ \rightarrow \pi^+ X$ amplitudes can
only be suppressed by a factor of a few. 

Another possibility raised
in the literature~\cite{Davoudiasl:2014kua} is that FCNCs could be
canceled between non-renormalizable and renormalizable (kinetic mixing)
couplings of the same vector. However, such tuning is only relevant for
$K^\pm \rightarrow \pi^\pm X$ decays (and even then can only reduce
the rate by a factor $\sim 10$);
$B$-physics is far less
affected by the kinetic mixing operators (and moreover $B \rightarrow K
X$ and $B \rightarrow K^* X$ decays are affected differently).

While we performed FCNC calculations for particular vector coupling models,
these conclusions apply more generally. Quark FCNC amplitudes
depend on the $XW^+W^-$ coupling (and the quark couplings) of the new
vector, so any models giving that coupling will lead to the similar decays.
Other couplings of $X$ to SM bosons have smaller effects;
a $XZZ$ vertex only contributes to less important low-energy processes,
while processes involving vertices of $ X $ with three SM gauge bosons are suppressed
by additional powers of the weak scale (or extra loop factors).
There can also be dimension-4 couplings involving more than one
$X$, such as an $XXW^+W^-$ vertex. However, if the $X$-SM couplings are all suppressed by some small parameter $g_X$,
then in most UV completions, $XX$-SM vertices will be suppressed by
$g_X^2$ limiting their capability to constrain these such couplings. 

For heavier vectors, exotic Higgs decays provide a relatively clean
probe, as long as the vector decays to SM states. We studied the decays
for both currents, finding that only for $ H ^\dagger D H $ can the
tree-level decays have a longitudinal enhancement, observable in $ h \rightarrow XX
$ decays. Nevertheless, approximate current conservation is broken at
one-loop leading to enhanced $ h \rightarrow ZX $ and $ h \rightarrow
XX $ decays for both currents. These conclusions rely on our assumption
that the theory does not have additional sources of electroweak symmetry
breaking other than the SM Higgs. If this is violated (e.g., by a generic
two Higgs doublet model), then enhanced decays can occur already at tree
level. Overall, we clarify how Higgs decay rates to $X$ can vary across
different models, and highlight their potential to probe vectors above
the few-GeV meson threshold.

\section{Acknowledgments:}
Research at Perimeter Institute is supported by the Government of Canada
through Industry Canada and by the Province of Ontario through the
Ministry of Economic Development \& Innovation. JD is supported in part
by the DOE under contract DE-AC02-05CH11231.


\bibliography{ZX}

\end{document}